\documentclass[slac_one]{revtex4}
\usepackage{graphicx}
\usepackage{fancyhdr}
\begin{document}

\title{Status of the ALICE experiment at the LHC} 

%

\author{P. Antonioli for the ALICE Collaboration}
\affiliation{INFN Bologna, Via Irnerio 46, 40126 Bologna, Italy}

\begin{abstract}
The status of the ALICE experiment is presented. ALICE
is the LHC experiment devoted to heavy ion collisions. Preparing
for the first lead-lead run, foreseen in November 2010, ALICE
is successfully collecting data in proton-proton collisions at 
the LHC since November 2009, exploiting the charateristics of the detector
for its proton-proton physics program. First results are briefly 
reviewed with an emphasis on performance of its detector sub-systems.
\end{abstract}

\maketitle

\thispagestyle{fancy}


\section{INTRODUCTION} 
ALICE~\cite{ppr1,ppr2} is a general-purpose heavy ion experiment designed to study the
physics of strongly interacting matter and the quark-gluon plasma
in nucleus-nucleus collisions at the LHC. Its physics program shaped
the detector design which has specific features compared to other LHC detectors.  
ALICE is actually designed to cope with the highest particle multiplicities for Pb-Pb 
collisions (at the design time a value of $dN_{ch}/dy$ up to 8000 was foreseen), 
providing excellent particle identification capabilities in a wide range 
of energies (0.1-100 GeV).

At the LHC the energy for nucleus-nucleus collision will be
30 times~\cite{Vogt} the highest reached at RHIC. This will in turn
correspond to a factor 2 in the temperature reached by the fireball
and an increase from 5-10 GeV/fm$^3$ up to 15-60 GeV/fm$^3$ in terms
of critical energy. The lifetime of the QGP will extend from
1.5-4.0 fm/c up to 10 fm/c. The study of the QGP is thus expected
to enter into a new domain, with many hard signals much more
available at the LHC energies (a factor 100 increase is
expected for $b\bar{b}$ production~\cite{bbcross}, for example).

As a main design guideline ALICE wants
to measure the flavor content and momentum phase distribution event-by-event.
Consequently ALICE has to track and identify most (within 2$\pi$$\cdot 1.8 \eta$ units) of the hadrons 
from very low momenta ($<$ 100 MeV/c) to study soft processess
up to fairly high $p_{T}$ ($\approx$ 100 GeV/c). The vertex recognition
of hyperons and D/B mesons must be operated in an environment of
very high charged-particle density. Dedicated and complementary
systems for di-electrons and di-muons and a devoted detector
to measure centrality must be in place. Photon detection and
electromagnetic calorimetry has been implemented in specific
regions of angular acceptance. 

The ALICE detector~\cite{AliceDet} can be broadly described by three groups of detectors: the central
barrel, the forward muon spectrometer (DIMUON) and the forward detectors. 
The central part of the detector measures hadrons, electrons 
and photons on an event-by-event basis. 
It covers polar angles from 45$^{o}$ to 135$^{o}$ over the full 
azimuth and it is embedded in the large L3 solenoidal magnet operated at B=0.5 T. The
central barrel consists of: an Inner
Tracking System (ITS) of high-resolution silicon detectors, a
cylindrical Time-Projection Chamber (TPC), a Transition Radiation Detector (TRD) for
electron identification, a Time-Of-Flight (TOF)
detector and a single-arm ring imaging Cherenkov detector (HMPID) 
for hadron identification. Two single-arm electromagnetic calorimeters (PHOS and EMCAL) complement
the characterization of each event. 
The forward muon arm (covering polar angles 180$^{o}$ - $\theta$ = 2$^o$
- 9$^o$) consists of a complex arrangement of absorbers, a large
dipole magnet, and fourteen planes of tracking and triggering
chambers. Several small detectors (ZDC, PMD, FMD, TZERO and VZERO) for global
event characterization and triggering are located at forward angles.
During the first LHC period of operations ALICE
has full hadron and muon capabilities, with the relevant detectors
fully installed and operational. TRD, PHOS and EMCAL detectors
have been partially installed and their completion is foreseen
during the first long LHC shutdown.

\begin{figure}[t]
\begin{minipage}[t]{0.48\linewidth}
\centering
\includegraphics[width=0.75\textwidth]{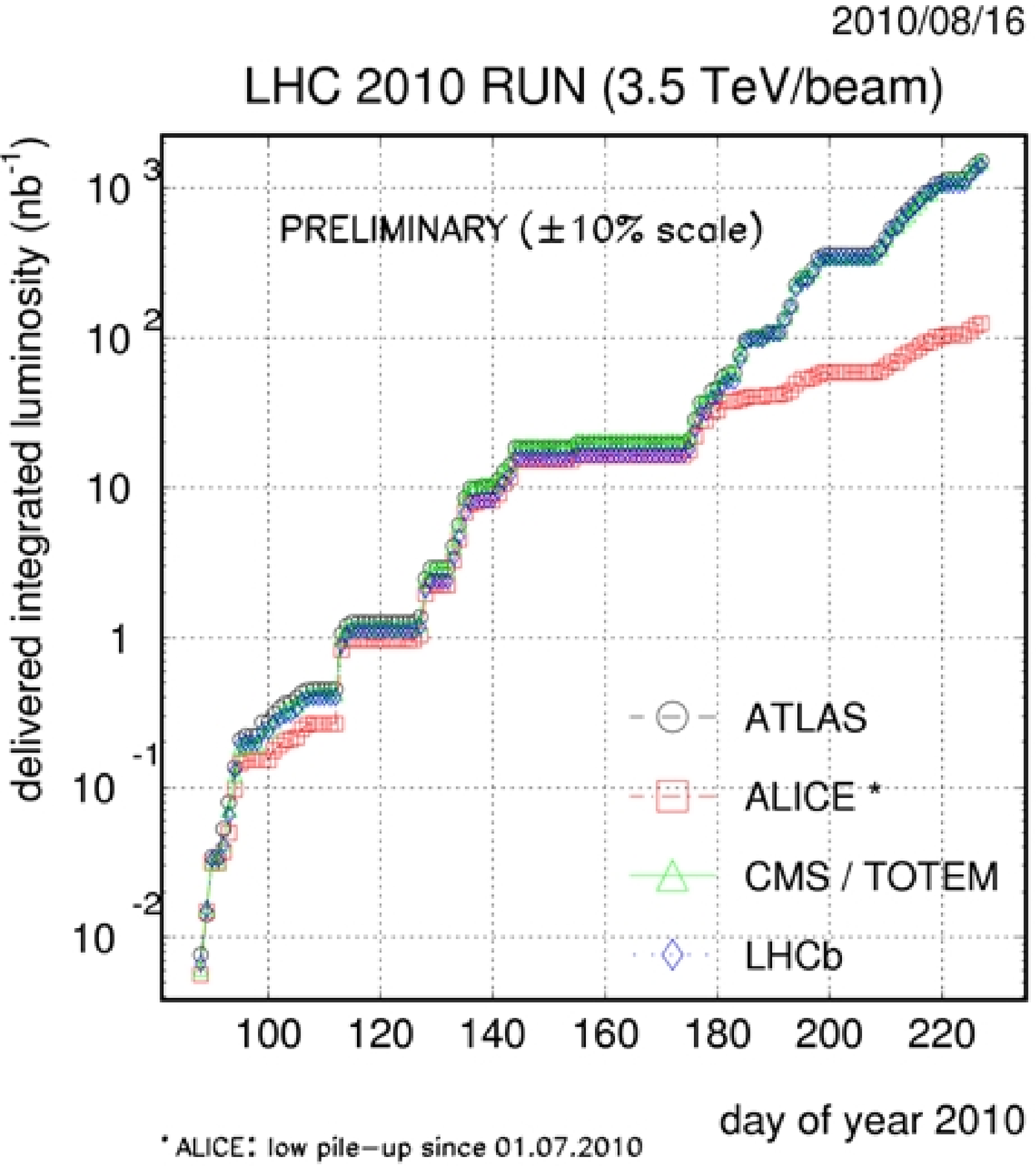}
\caption{The integrated luminosity delivered at the four LHC experiments as of 16 August 2010. 
Note ALICE low pile-up settings since beginning of July.}
\label{fig:lumi}
\end{minipage}
\hfill
\begin{minipage}[t]{0.48\linewidth}
\centering
\includegraphics[width=0.75\textwidth]{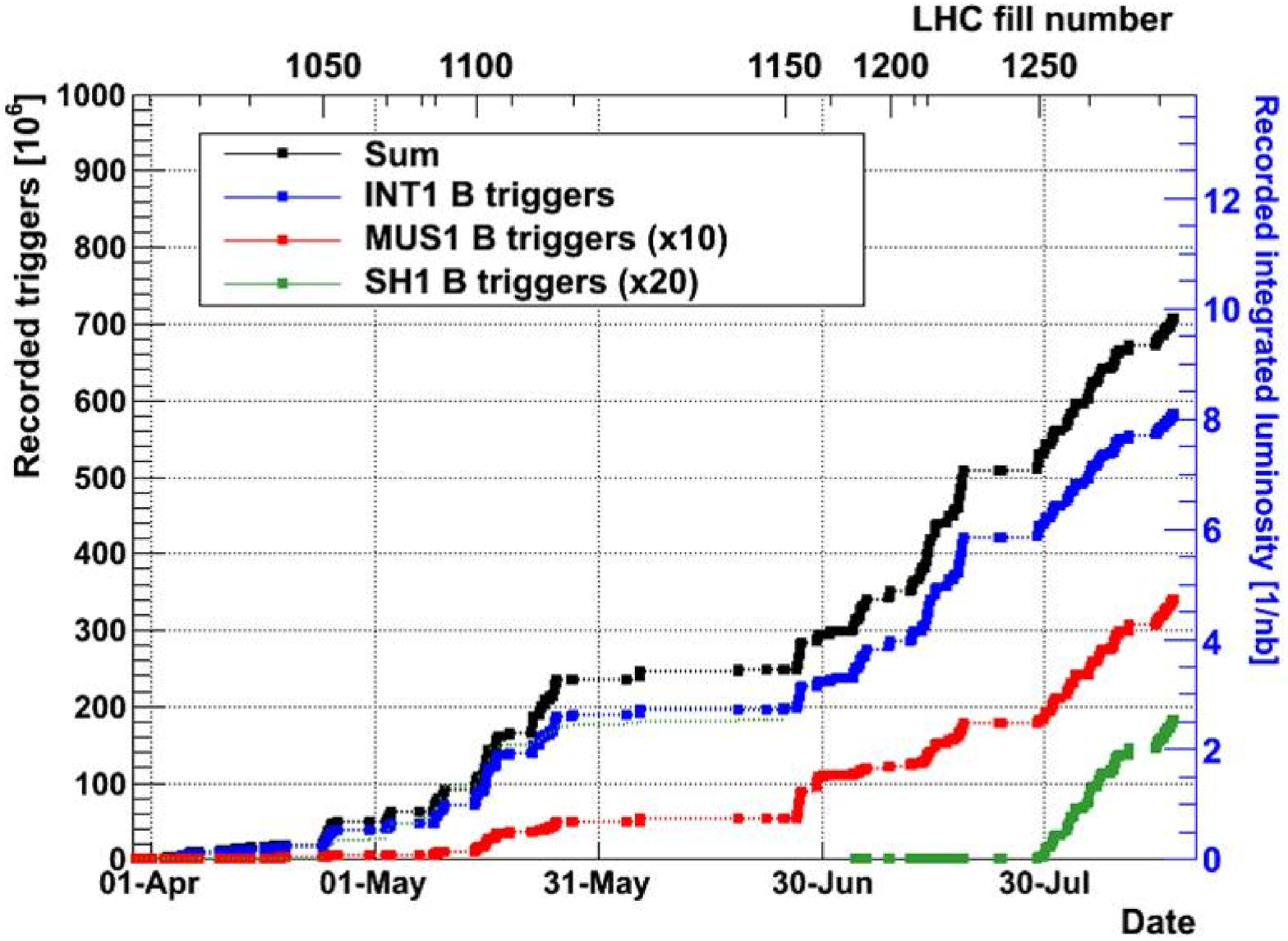}
\caption{Data collection in ALICE for the different trigger types.}
\label{fig:triggers}
\end{minipage}
\end{figure}

Data collection is proceeding smoothly thanks also to the
excellent performance of the LHC accelerator. Preparing for lead-lead collisions
ALICE aims to collect in 2010 1 billion of proton-proton minimum bias (MB)
events. Differently from hard processess,
the particle production in soft interactions is not well-established
in QCD and phenomenological models are used.
Even at LHC energies a significant fraction of the produced particles
doesn't originate from hard scattering processes 
and it is therefore needed to collect data in order 
to tune these models and provide a solid baseline picture to then
study heavy ion collisions. ALICE, with its unique particle identification
capabilities and low momentum reach, can in particular complement
similar studies made by other LHC experiments during the early
data collection at the LHC.

The MB trigger requires a hit in the SPD or in either one of
the VZERO counters; i.e. essentially at least one charged
particle anywhere in 8 units of pseudorapidity. The MB trigger
is complemented by the coincidence with the signals from two beam pick-up
counters, one on each side of the interaction region.
The SPD (the ITS innermost detector) surrounds the central beryllium
beam pipe (3 cm radius, 0.23\% of a radiation length) with
two cylindrical layers (at radii of 3.9 and 7.6 cm, 2.3\% of
a radiation length) and covers the pseudorapidity ranges
$\eta < 2$ and $\eta < 1.4$ for the inner and outer layers, respectively. 
VZERO counters consist of two scintillator hodoscopes.
They are placed on either side of the interaction region at z = 3.3 m and z = -0.9 m.
They cover the regions 2.8 $< \eta <$ 5.1 and -3.7 $<\eta<$ -1.7.

The integrated luminosity delivered to the four LHC experiments is shown in
Fig.~\ref{fig:lumi}. It is worthwhile to note that since beginning of July 2010
the luminosity at ALICE point of interaction has been reduced
to keep the pile-up within the TPC less than 5\%. 
The data collected so far for the different type of triggers
are shown in Fig.~\ref{fig:triggers}. Besides the discussed MB trigger (INT1B),
a devoted trigger for the di-muon spectrometer (MUS1B) is in place. It requires,
in addition to the MB trigger, at least 1 hit in three of the four triggering 
layers of the di-muon detector with some geometrical alignment request, corresponding
currently to an effective cut of $\sim$ 0.5 GeV/c in transverse momentum
of the selected tracks. Since the beginning of August a high-multiplicity
trigger (SH1B), again based on the SPD and requiring at least 65 hits firing in the detector,
has been activated.  

Section 2 of this paper highlights some key performances results. 
Section 3 presents some early physics results but with an emphasis on the detector 
capabilities which made them possible. A full overview of the
first physics results is instead presented at this conference in~\cite{jfh}
and the ALICE physics program for heavy ions in~\cite{cp}.

\begin{figure}[t]
\begin{minipage}[t]{0.48\linewidth}
\centering
\includegraphics[width=0.75\textwidth]{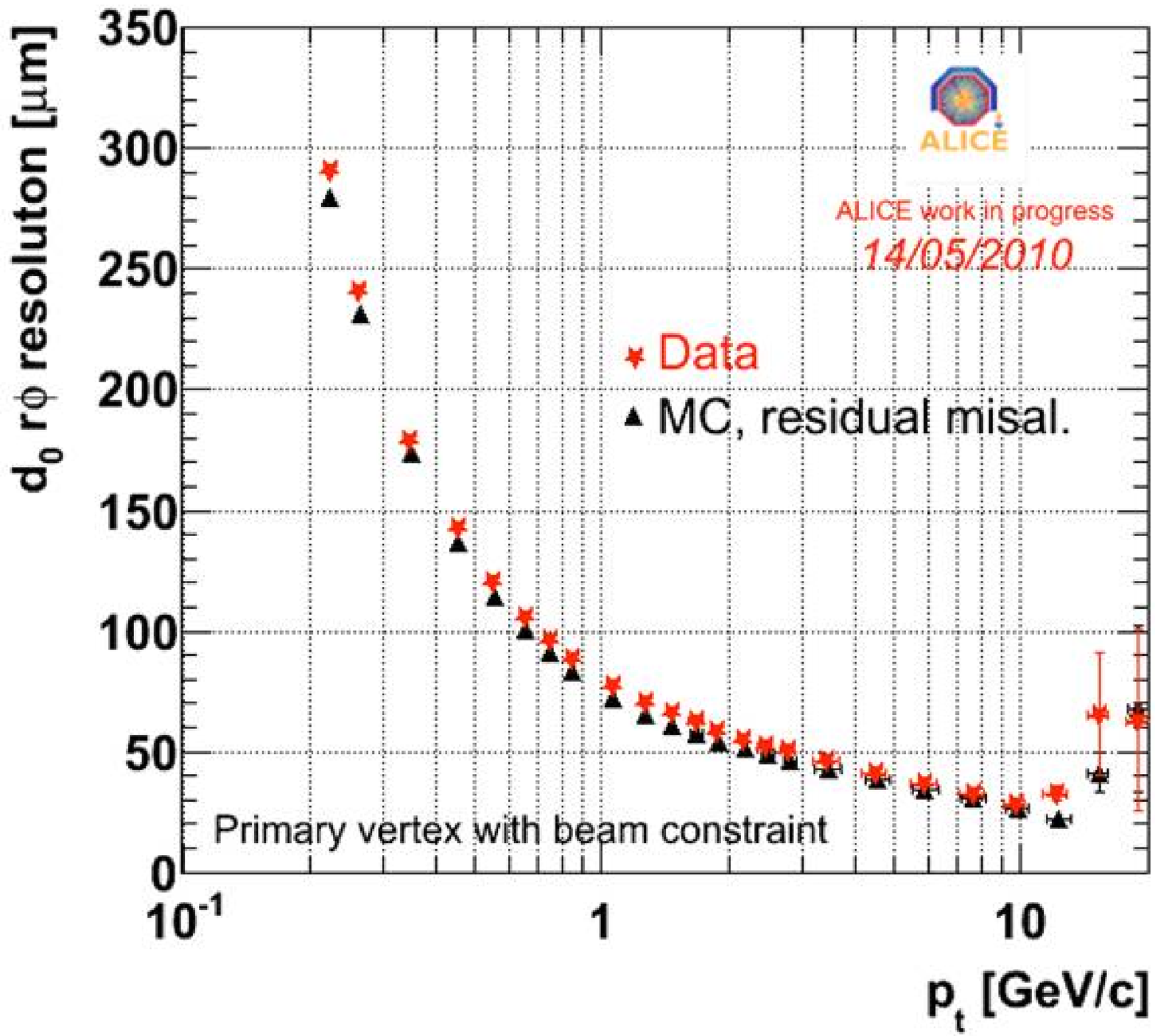}
\caption{Impact parameter resolution in the SPD as a function of transverse momemtum.}
\label{fig:impact}
\end{minipage}
\hfill
\begin{minipage}[t]{0.48\linewidth}
\centering
\includegraphics[width=0.95\textwidth]{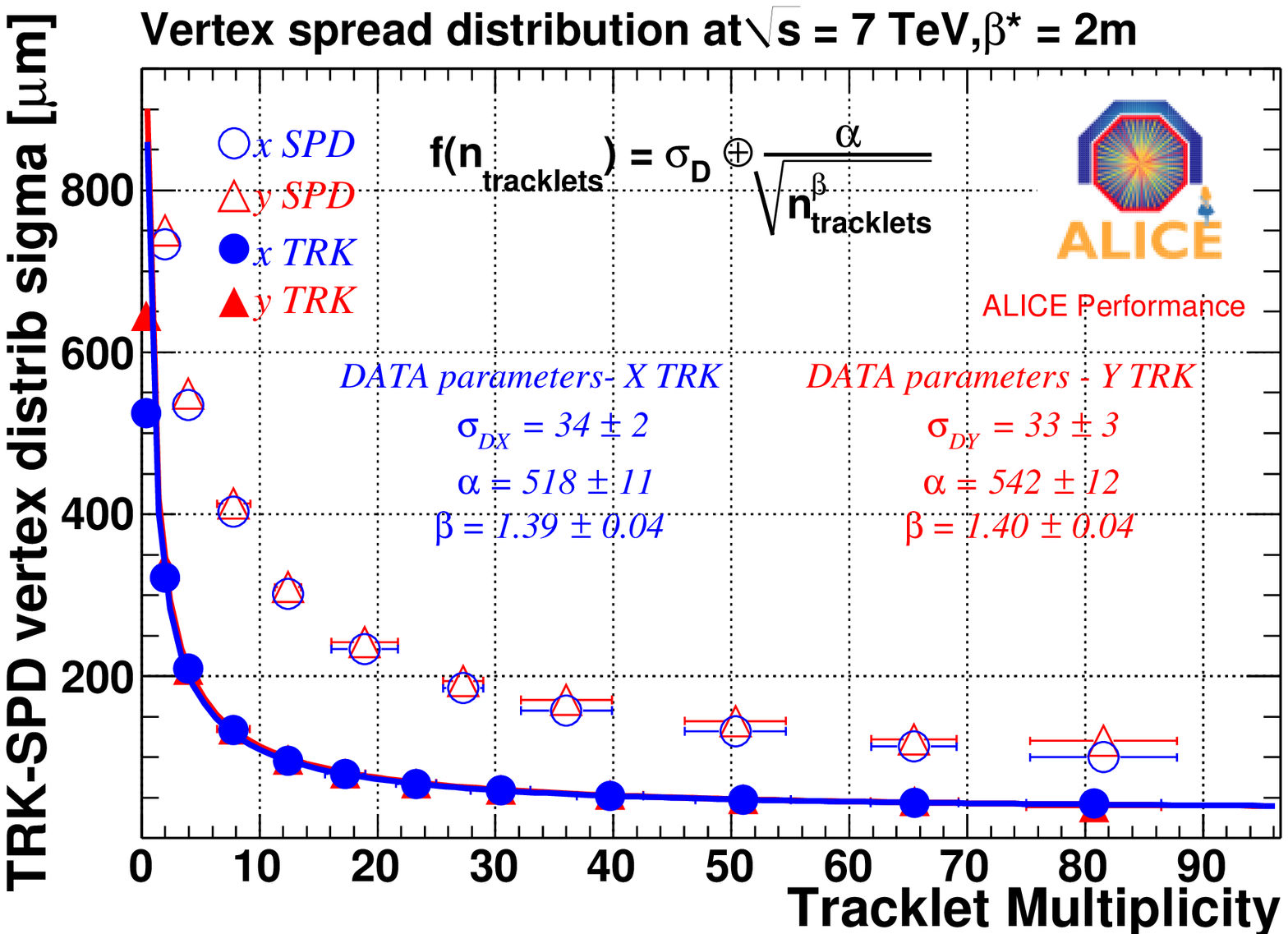}
\caption{Vertex spread measured by ITS (SPD) and ITS and TPC (TRK) reconstructed tracks.}
\label{fig:vertex}
\end{minipage}
\end{figure}

\begin{figure}[h]
\begin{minipage}[t]{0.48\linewidth}
\centering
\includegraphics[width=0.75\textwidth]{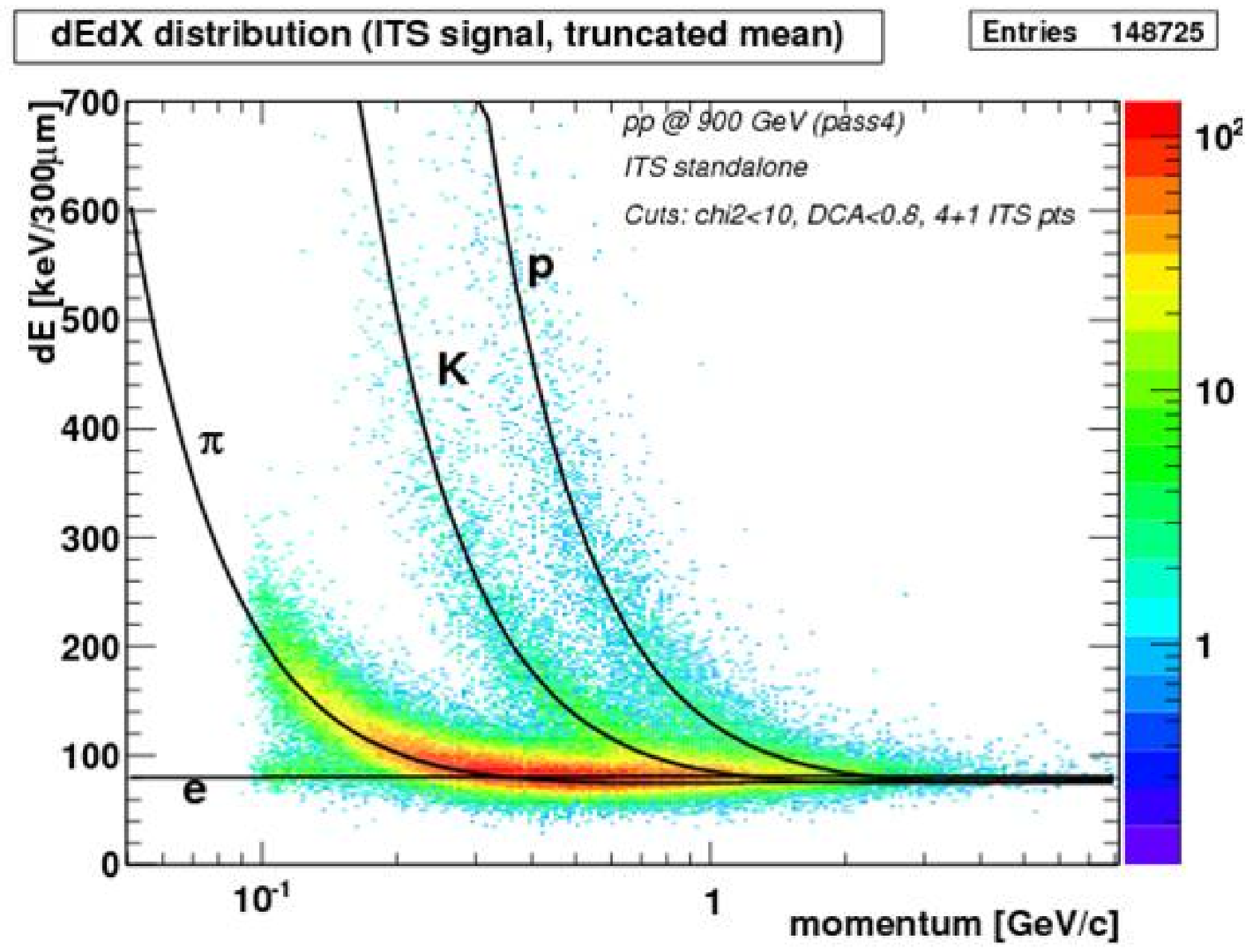}
\caption{Particle Identification provided by ITS via dE/dx measurement.}
\label{fig:pidits}
\end{minipage}
\hfill
\begin{minipage}[t]{0.48\linewidth}
\centering
\includegraphics[width=0.75\textwidth]{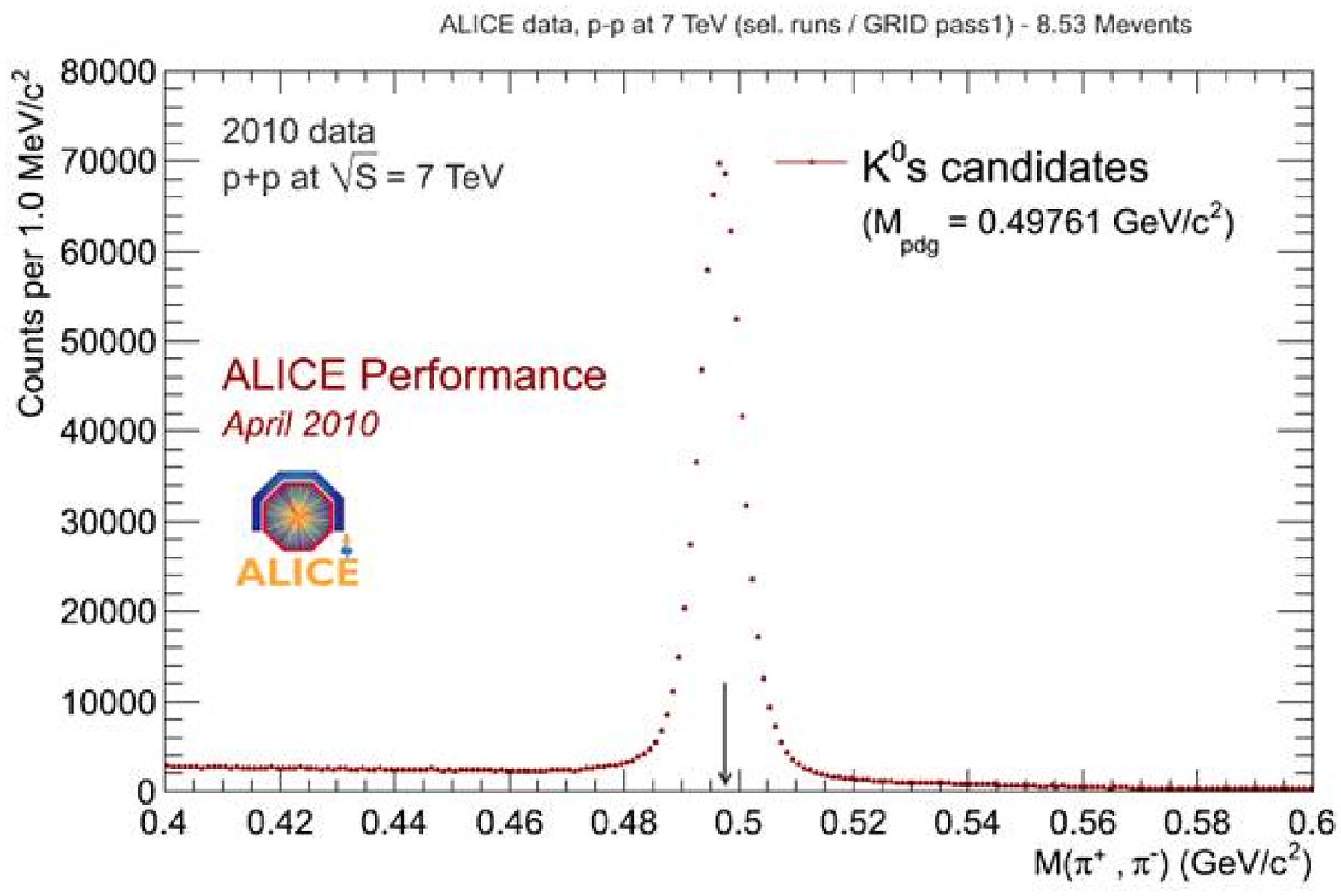}
\caption{$K^0 _s$ invariant mass.}
\label{fig:k0mass}
\end{minipage}
\end{figure}

\section{PERFORMANCE}


The Inner Tracking System (ITS) of ALICE consists of six silicon layers, the two innermost 
abovementioned pixel detectors (SPD), two layers of drift detectors (SDD) and two outer layers
of strip detectors (SSD), which provide up to six points for each track. The design spatial resolutions
of the ITS sub-detectors ($\sigma _{r\phi} \times \sigma _z$ are:  12 $\times$ 100 $\mu$m$^2$ (SPD), 
35 $\times$ 25 $\mu$m$^2$ (SDD) and 20 $\times$ 830 $\mu$m$^2$ (SSD). 

Fig.~\ref{fig:impact} shows the impact parameter resolution for reconstructed tracks
in ALICE satisfying standard TPC track quality cuts (basically a number of TPC clusters greater than 70)
and having two points in the SPD.  For each track, its impact parameter was estimated 
with respect to the primary vertex reconstructed without using this track.  The resulting resolution is the 
convolution of the track-position and the primary-vertex resolutions. While the difference with the Monte 
Carlo shows that the residual misalignment
is already below 10 $\mu$m, the resolution of 50 $\mu$m achieved at 2 GeV/c is critical.
One of the main items of the ALICE physics program is the measurement of charm and beauty hadron 
production in Pb-Pb collisions. To  measure the separation from the interaction vertex, of the 
decay vertices of heavy flavoured hadrons,
requires a resolution on the distance of the closest approach to the vertex well below 100 $\mu$m.

The spread of the measured vertex position has been studied similarly using tracks reconstructed
both in the TPC and the SPD and with the SPD only. The result is presented in Fig.~\ref{fig:vertex}.
The asymptotic limit estimates the size of the luminous region ($\sim$ 30 $\mu$m for both projections) 
seen for the vertices reconstructed with tracks. The vertex resolution depends on the event multiplicity.
It can be parametrized as 540 $\mu m/(N_{SPD})^{0.45}$ in x and y and 550 $\mu m/(N_{SPD})^{0.6}$ in z,
where N$_{SPD}$ corresponds to the number of SPD tracklets (a tracklet is built by associating pairs of 
hits in the two layers of the SPD). These results show the ITS is operating close to 
its design parameters. 
The ITS via measurement of $dE/dx$ in  its strip and drift detectors provides particle identification. ALICE 
can therefore identify particles at very low momentum ($<$ 500 MeV) as shown in Fig.~\ref{fig:pidits}.

The TPC, 5 m long, is used to record charged particle tracks as they leave
ionization trails in the Ne-CO$_2$-N$_2$ gas. The ionization electrons drift up to 2.5 m to be measured
on 159 pad rows. The position resolution is better than 2 mm. At the present level of calibration, the transverse
momentum resolution achieved in the TPC is given by $(\sigma (p_T)/p_T)^2 = (0.01)^2 + (0.007 \cdot p_T)^2$, with $p_T$
in GeV/c. For $p_T > 1$ GeV/c the resolution was measured in cosmic muon events (comparing the momentum measured
in the upper and lower halves of the TPC), for $p_T < 1$ GeV/c the Monte Carlo estimate of $\sigma (p_T)/p_T \approx$ 1$\%$ 
has been cross-checked using the $K ^0 _s$ invariant mass distribution which is shown in Fig.~\ref{fig:k0mass}. 
The calibration of the absolute momentum scale was further verified using the invariant
 mass spectra of $\Lambda$, $\bar{\Lambda}$, $K^0 _s$ and $\phi$. The reconstructed peak positions agree 
with their PDG values within 0.3 MeV/c$^2$.  

\begin{figure}[t]
\begin{minipage}[t]{0.48\linewidth}
\centering
\includegraphics[width=0.75\textwidth]{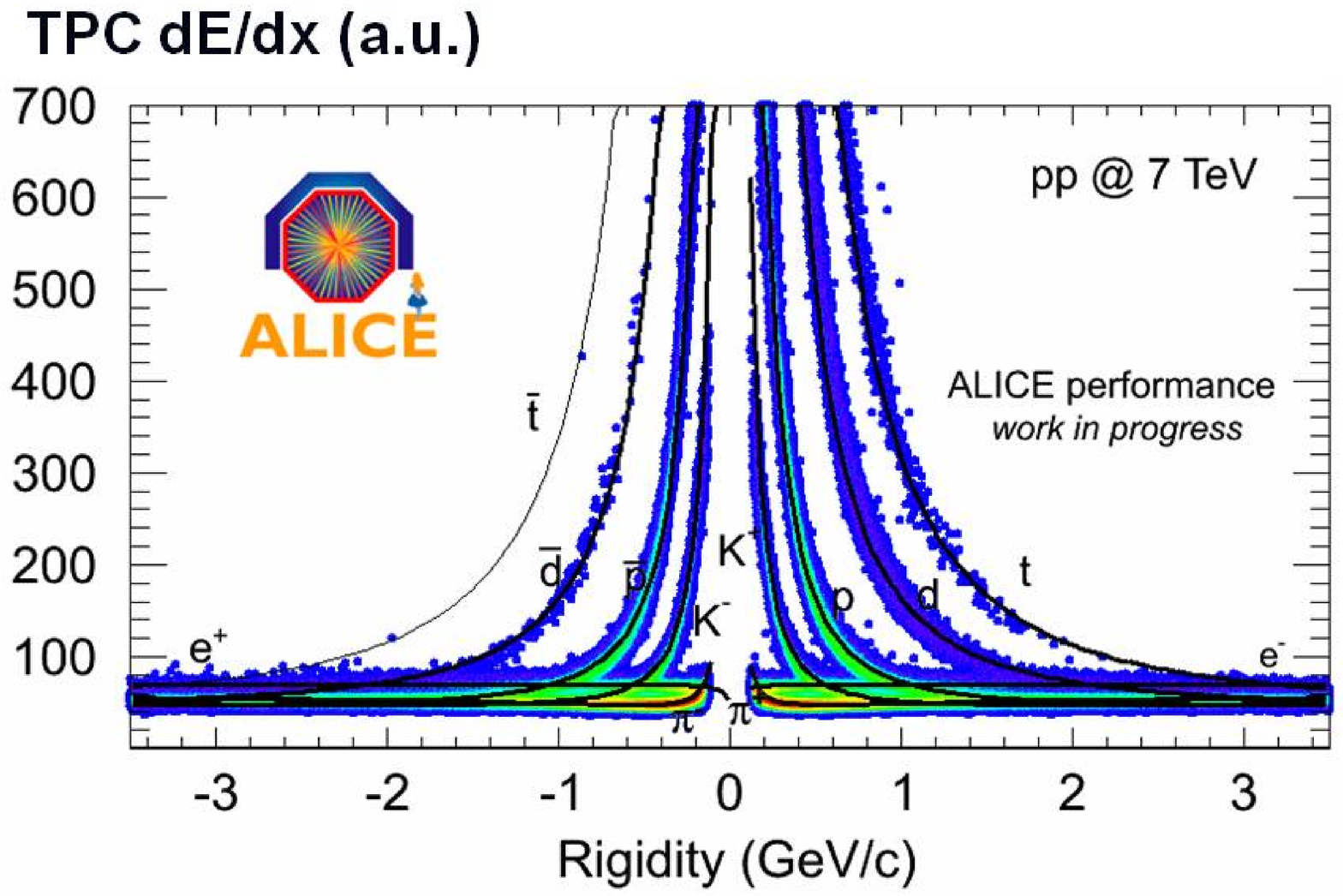}
\caption{Particle Identification provided by TPC via dE/dx measurement.}
\label{fig:pidtpc}
\end{minipage}
\hfill
\begin{minipage}[t]{0.48\linewidth}
\centering
\includegraphics[width=0.75\textwidth]{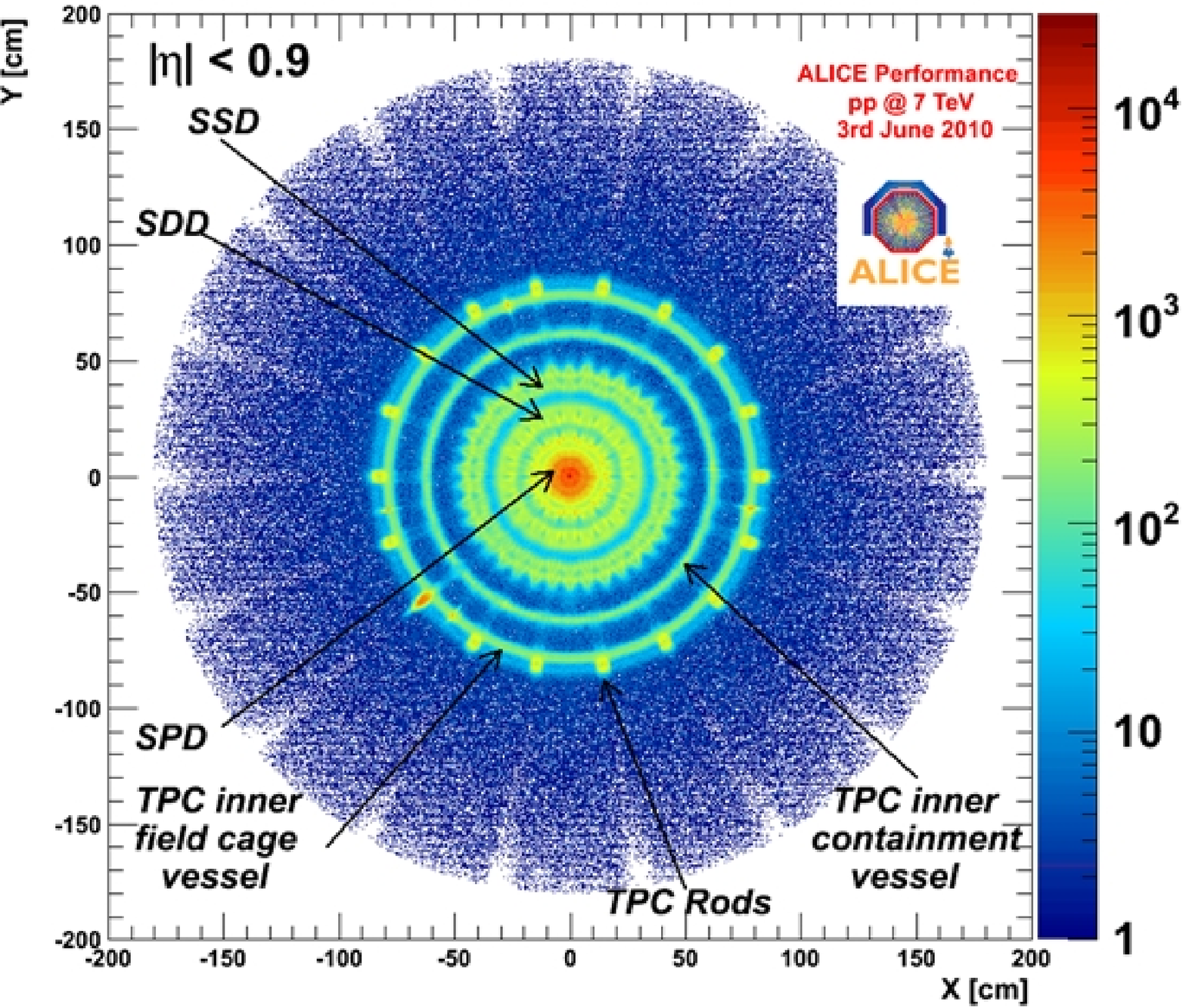}
\caption{Reconstructed vertices of gamma conversions, tracking $e^+ e^-$ pairs with the TPC.}
\label{fig:gamma}
\end{minipage}
\end{figure}

The measurement of dE/dx and rigidity allows the TPC to identify $\pi$, $K$ and p at intermediate transverse
momenta. The separation achieved by the TPC is shown in Fig.~\ref{fig:pidtpc}: despite
the relatively low statistics it is wortwhile to highlight the clear identification of deuterons and tritium nuclei (and their
anti-particles). This opens up the possibility for ALICE at LHC to successfully study in nuclei-nuclei collisions the formation of
exotic  hypernuclei, as reported this year by STAR at RHIC~\cite{star}.

\begin{figure}[t]
\begin{minipage}[t]{0.48\linewidth}
\centering
\includegraphics[width=0.75\textwidth]{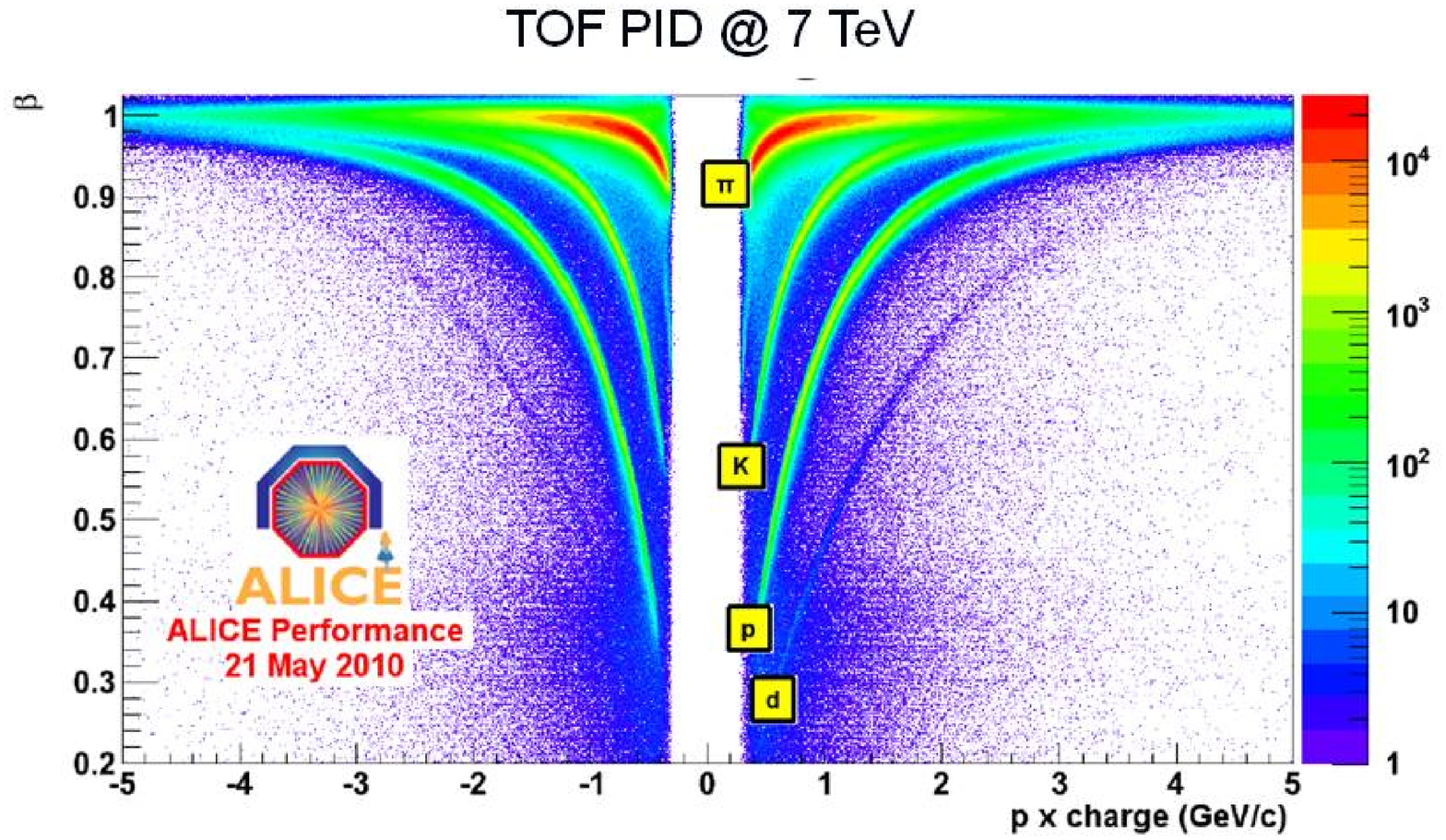}
\caption{Particle identification provided by TOF via $\beta$ measurement.}
\label{fig:pidtof}
\end{minipage}
\hfill
\begin{minipage}[t]{0.48\linewidth}
\centering
\includegraphics[width=0.75\textwidth]{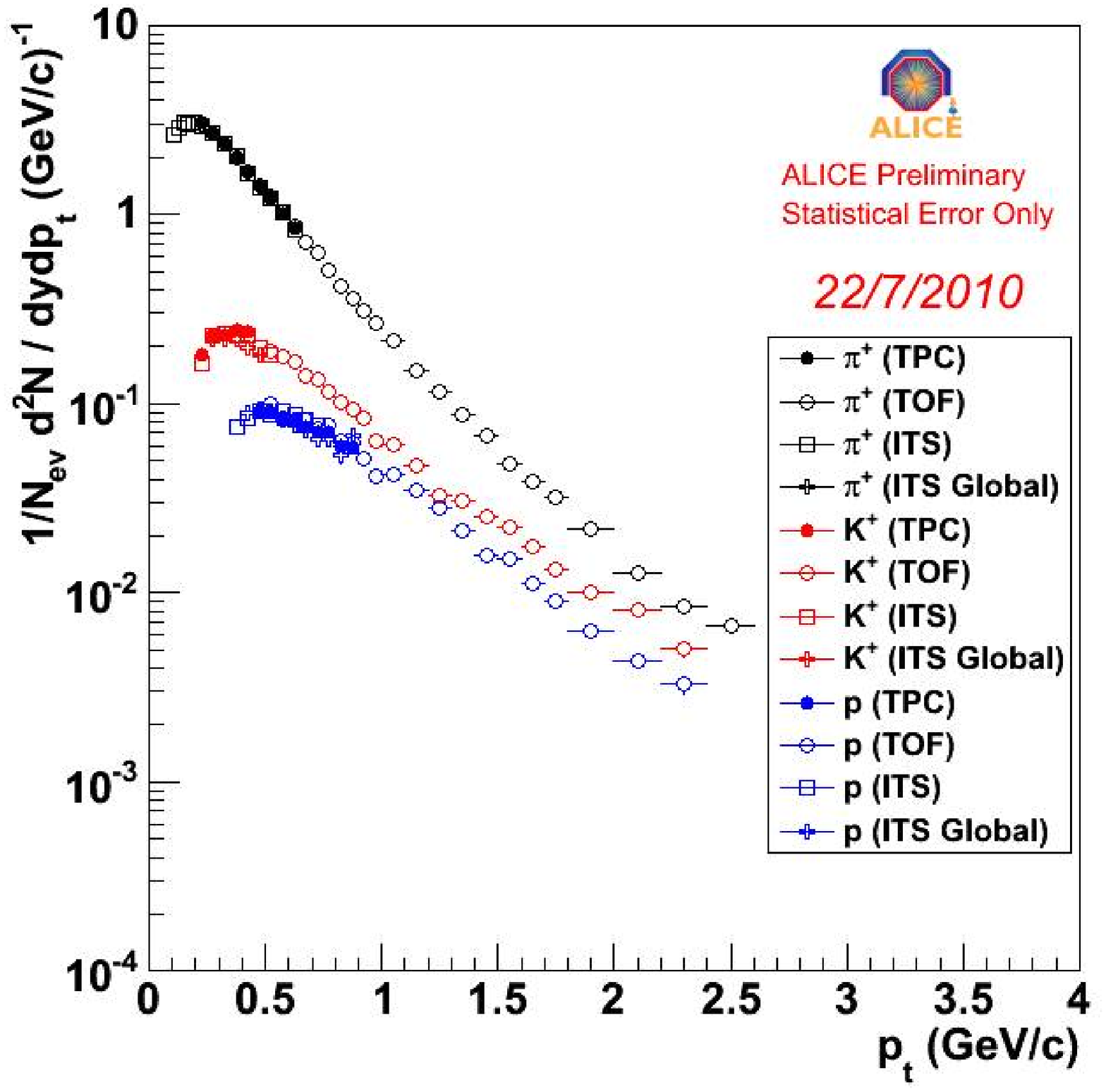}
\caption{Identified spectra of hadrons, using ITS, TPC and TOF information for pp collisions at $\sqrt{s}=$900 GeV.}
\label{fig:spectra}
\end{minipage}
\end{figure}

The material budget of the detector has been mapped measuring the vertex of
photon conversions as tracked by the TPC. The amount of material in the central part of
ALICE is very low, corresponding to about 10$\%$ of a radiation length on average between the vertex
and the active volume of the TPC. Such pattern is clearly visibile in Fig.~\ref{fig:gamma} where
the three different silicon detectors, the segmentation and the support structure of the TPC are precisely
identified. The current simulation reproduces the amount and the spatial distribution of
reconstructed conversion points in great detail, with a relative accuracy of a few percent. Such knowledge
of the detector has been key to keep under control systematic uncertainties while measuring
the $\bar{p}/p$ ratio. 

The particle identification capabilities of ALICE are extended at higher momentum (up to 2.5 GeV/c for pions and kaons
and up to 4 GeV/c for protons) by the TOF detector, a large (144 m$^2$) 
Multigap RPC array with resolution below 100 ps at current level of calibration.
Its performance is shown in Fig.~\ref{fig:pidtof}, where the separation between the different hadrons
is clearly visible, including deuterons. Preliminary spectra of identified positive hadrons,
using the information from ITS, TPC and TOF are shown in Fig.~\ref{fig:spectra} for data taken at $\sqrt{s}=900$ GeV. 
Despite it is still a preliminary result, this plot shows neatly the good understanding of sub-detectors and inter-calibration issues. 
Other cross-checks have been exploited using different techniques to measure kaons spectra: charged kaons directly identified  (using time of flight or dE/dx measurement) or via detection of their kink-decay topology and 
decay vertex reconstruction of $K^0 _s$.
Especially using
the data collected at $\sqrt{s}=7$ TeV, identified spectra are expected to provide a baseline reference for
the forthcoming measurements in lead-lead collisions.

\section{FIRST PHYSICS RESULTS AND CONCLUSIONS}
\begin{figure}[t]
\begin{minipage}[t]{0.48\linewidth}
\centering
\includegraphics[width=0.75\textwidth]{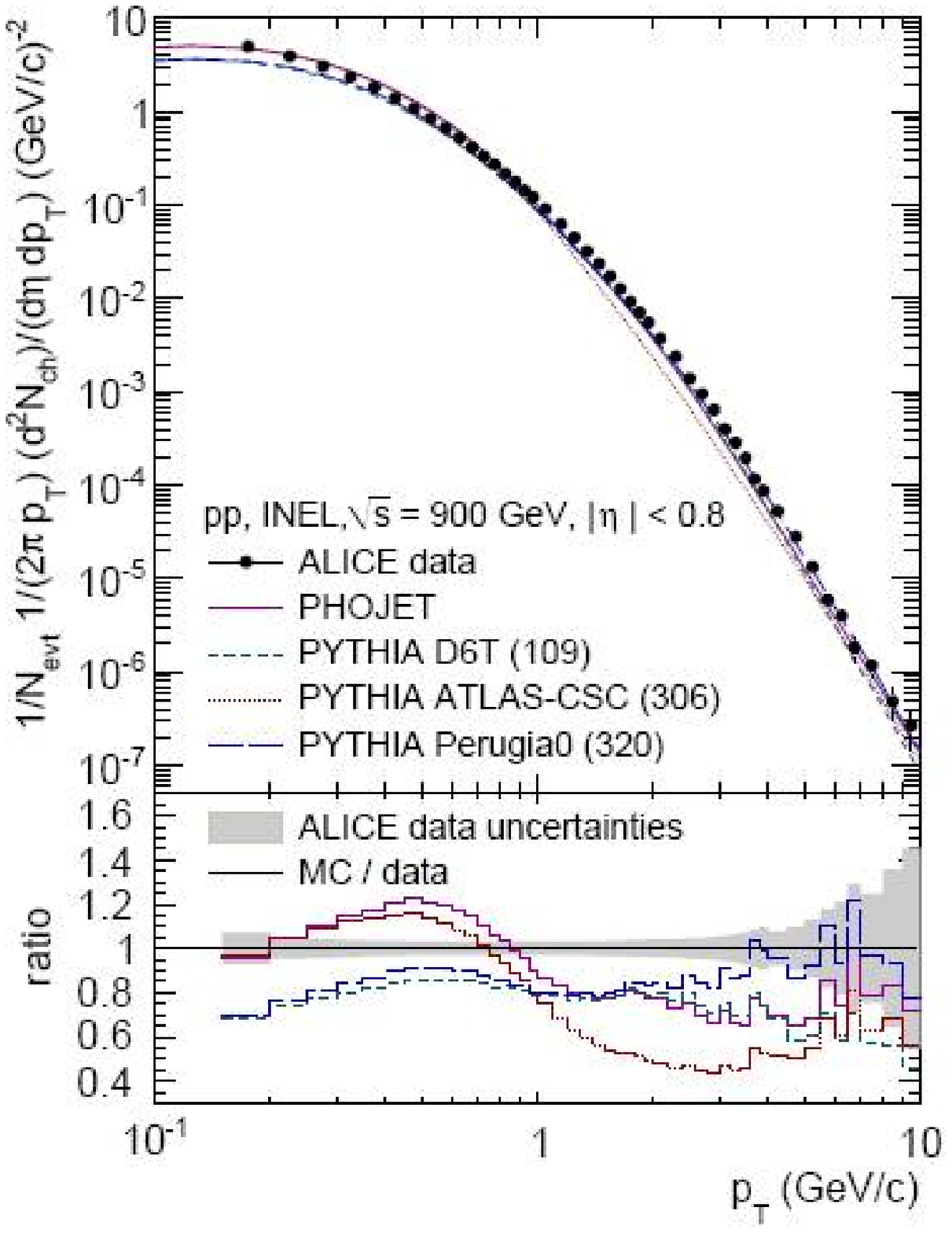}
\caption{Charged particle differential yield compared with various models. Bottom: ratio
Monte Carlo/data.}
\label{fig:spectra09mc}
\end{minipage}
\hfill
\begin{minipage}[t]{0.48\linewidth}
\centering
\includegraphics[width=0.75\textwidth]{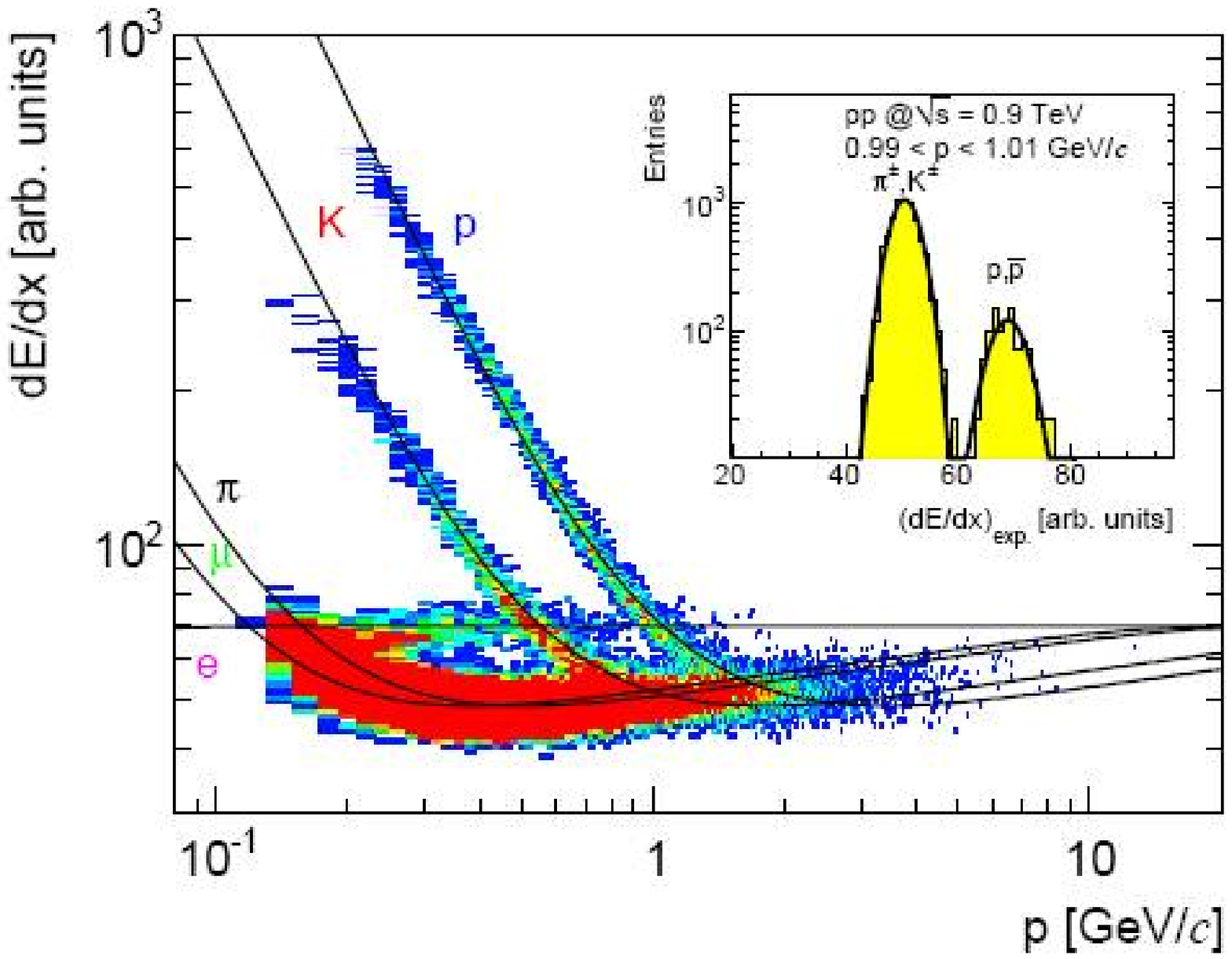}
\caption{Measured ionization per unit length in the TPC for $\bar{p}$/p ratio analysis (see text). }
\label{fig:pratiodedx}
\end{minipage}
\end{figure}

ALICE has already published physics results about charged particle 
multiplicity at $\sqrt{s}$=0.9~\cite{mult09}, 2.36~\cite{mult236}
and 7~\cite{mult7} TeV. The ITS and TPC detectors, coupled with the
discussed MB trigger, are at the basis of these results.
The measured value at the highest energies are significantly
higher than that obtained from current models and tunes. This
fact may be interpreted as a good news for heavy ions physics:  
at LHC we could expect a hotter and dense medium in lead-lead
collisions. Transverse momentum spectra of charged
particles at $\sqrt{s}=0.9$ TeV~\cite{spectra09} shown in Fig.~\ref{fig:spectra09mc} 
provided further insights about required Monte Carlo tuning: while
it is possible to change some parameters to obtain one or two of
the observed features it is more complicated to reproduce
all observed characteristics. The full set of ALICE observables (including identified spectra 
for $K ^0 _s$, $\Lambda$, $\Xi$, $\phi$) not only will help to tune Monte Carlo
parameters, but will shed hopefully further
light on our understanding of soft QCD.

\begin{figure}[t]
\begin{minipage}[t]{0.48\linewidth}
\centering
\includegraphics[width=0.75\textwidth]{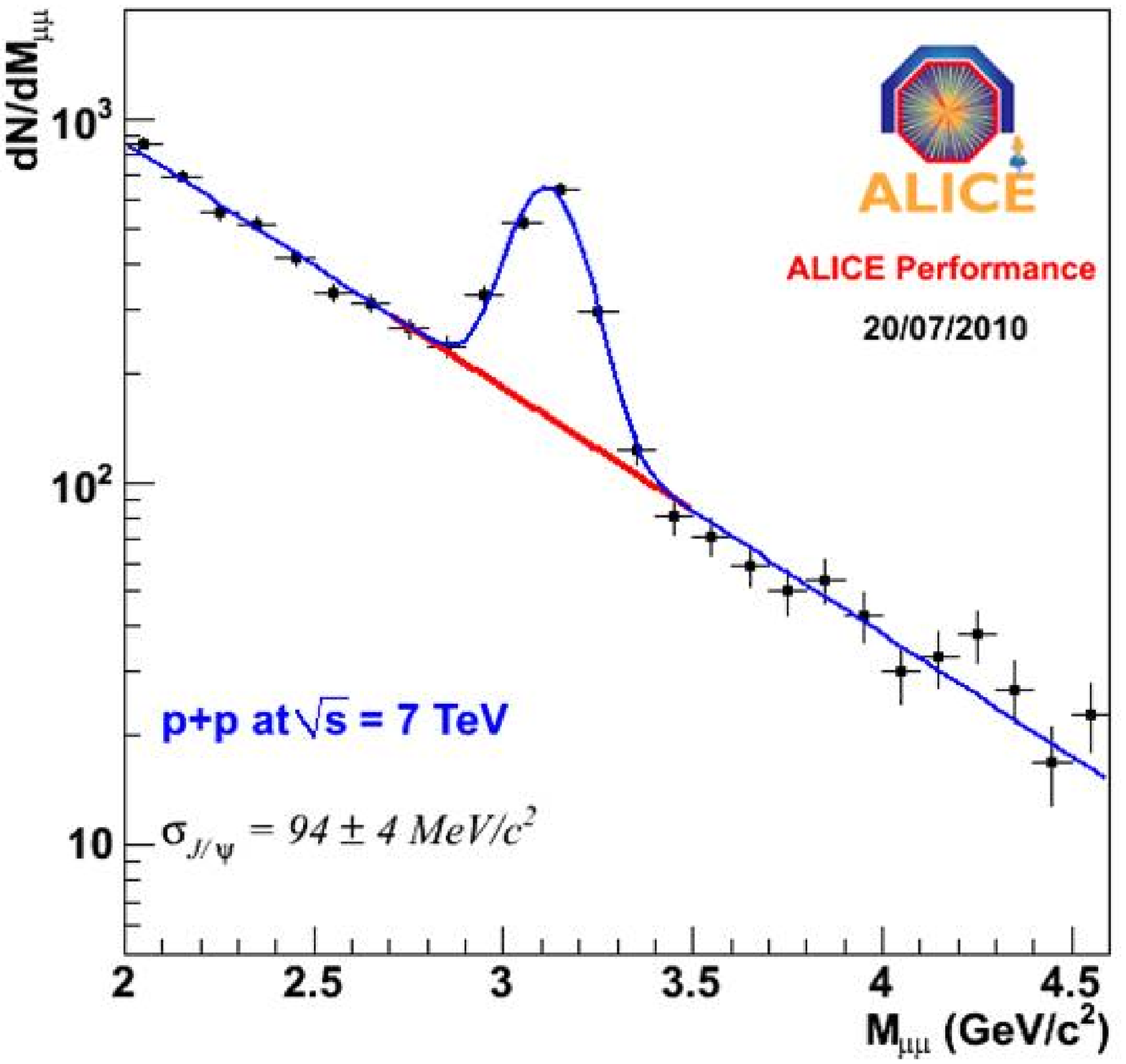}
\caption{J/$\Psi$ reconstructed mass with muon pairs selected in the forward muon spectrometer.}
\label{fig:jpsi}
\end{minipage}
\hfill
\begin{minipage}[t]{0.48\linewidth}
\centering
\includegraphics[width=0.75\textwidth]{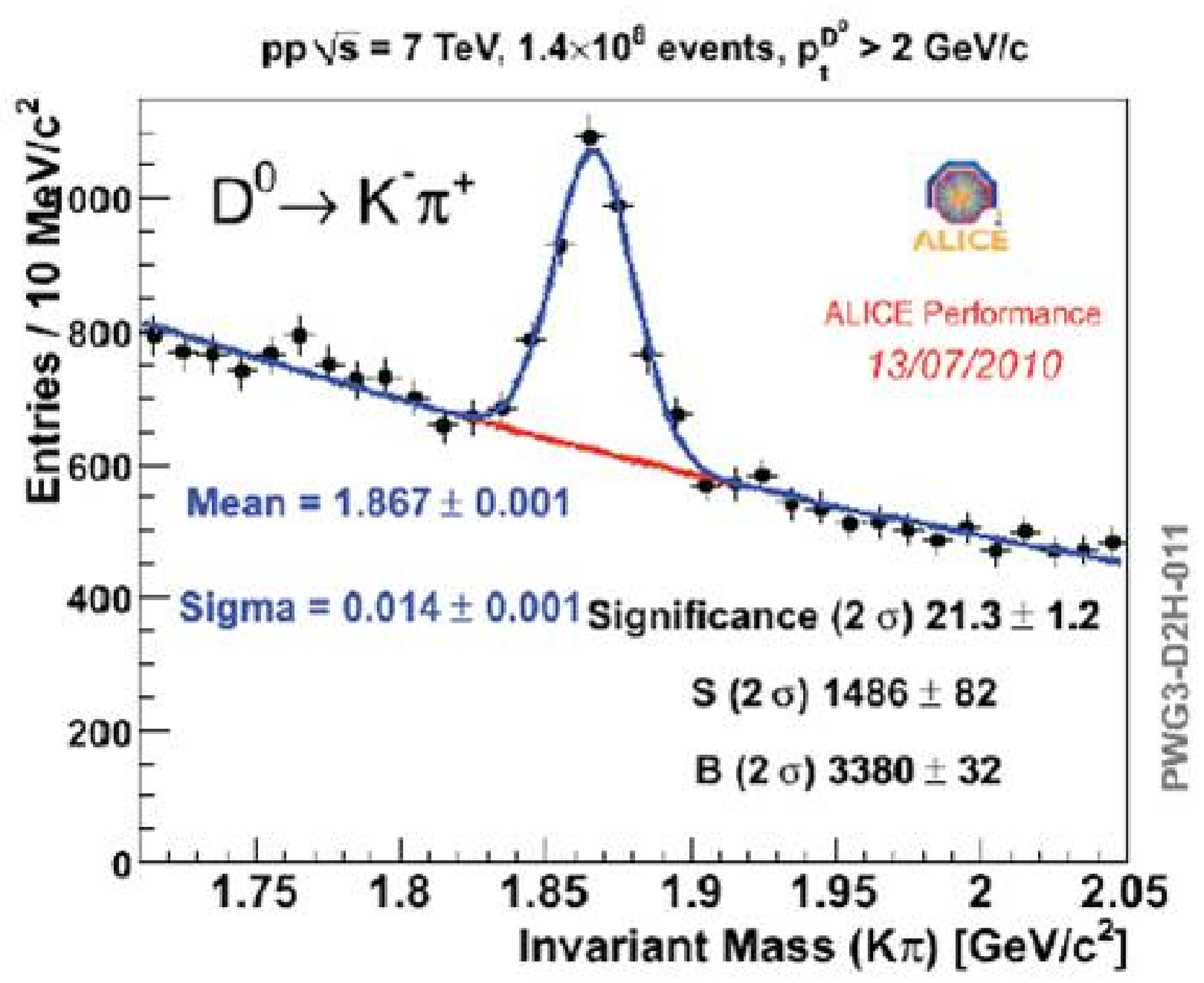}
\caption{D$^0$ mass reconstructed in the barrel using ITS, TPC and TOF information.}
\label{fig:dbarrel}
\end{minipage}
\end{figure}

ALICE studied also the $\bar{p}$/p ratio~\cite{pratio} at 
$\sqrt{s}$=0.9 and 7 TeV at mid-rapidity and 0.45 $< p_T <$ 1.05 GeV/c.
Its results set tight limits on additional contributions to baryon-number transfer
over very large rapidity intervals with
respect to standard models. Experimentally, besides the 
precise mapping of material budget already discussed, this result is grounded
on the excellent PID capabilities of the TPC in the specified
momentum range, as shown in Fig.~\ref{fig:pratiodedx} (the inset shows the measured
ionization for tracks with 0.99$<p<$1.01 GeV/c with Gaussian fits to the data).

Discussed tracking, momentum and PID performances are also
at the basis of preliminary results related to J/$\Psi$ spectra
(using the forward muon spectrometer) and identification of open-charm mesons decays
in the barrel shown respectively in Fig.~\ref{fig:jpsi}
and~\ref{fig:dbarrel}. The alignment procedure 
played also a key role: the width of the J/$\Psi$ peak moved
from an initial value of 240 MeV/c$^2$ to 95 MeV/c$^2$ with a target design resolution of 70 MeV/c$^2$.

In conclusion, during the first months of operation of the CERN-LHC, 
ALICE is taking data smoothly and with all its detectors
in operation. At the time of the conference more than 700 millions of triggers were
taken, with a recorded integrated luminosity of 10 nb$^{-1}$.
Real collisions allowed  to check if ALICE 'fits for purpose'. 
As shown, most sub-detectors are  at or close to design parameters. 
The calibration, intercalibration
and understanding of different detectors is in continuous progress.

The ALICE detector is now eagerly waiting for heavy ion
collisions: the energy reached at LHC will allow the study
of the Quark Gluon Plasma in a new domain.
The physics analysis for proton-proton collisions is well
under way with results already published for multiplicity,
Bose-Einstein correlations~\cite{bec}, p$_T$ spectra, $\bar{p}$/p ratio
and a wide range of analyses is on-going. The unique PID capabilities
and low p$_T$ reach in ALICE are being exploited
in forthcoming analyses.


\vspace{-0.5cm}
\begin{thebibliography}{9}   
\bibitem{ppr1}
ALICE Collaboration, J. Phys. G, 30, 1517 (2004).
\bibitem{ppr2} 
ALICE Collaboration, J. Phys. G, 32, 1295 (2005).
\bibitem{Vogt}
R. Vogt, Nucl. Phys. A752, 447 (2005)
\bibitem{bbcross}
M. Mangano, R. Vogt and S. Frixione (editors), CERN Yellow Report, CERN-2004-009, 247 (2004).
\bibitem{AliceDet}
ALICE Collaboration, J. Instr. 3, S08002 (2008).
\bibitem{jfh}
Jan Fiete Grosse-Oetringhaus for the ALICE Collaboration, these proceedings (2010).
\bibitem{cp}
Claude Pruneau for the ALICE Collaboration, these proceedings (2010).
\bibitem{star}
STAR Collaboration, Science, 328, 58 (2010). 
\bibitem{mult09}
ALICE Collaboration, Eur. Phys. J. C, 65, 111 (2010).
\bibitem{mult236}
ALICE Collaboration, Eur. Phys. J. C, 68, 89 (2010).
\bibitem{mult7}
ALICE Collaboration, Eur. Phys. J. C, 68, 345 (2010).
\bibitem{spectra09}
ALICE Collaboration, Phys. Lett. B, 693, 53 (2010).
\bibitem{pratio}
ALICE Collaboration, Phys. Rev. Lett., 105, 072002 (2010).
\bibitem{bec}
ALICE Collaboration, Phys. Rev. D, 82, 52001 (2010).

\end{thebibliography}
\end{document}